\documentclass[12pt]{article}
\usepackage{fullpage,epsfig,fancyheadings}
\usepackage[psamsfonts]{euscript}
\usepackage{epic}
\begin{document}
\vspace*{-.6in}
\begin{flushright}
CALT-68-2414\\
DOE RESEARCH AND\\
DEVELOPMENT REPORT
\end{flushright} 

\vspace{.3in}
{\Large
\begin{center}
{\bf Neutrino Mass Matrix and Hierarchy}  
\end{center}}
\vspace{.2in}

\begin{center}
Peter Kaus\\ 
Physics Department, University of California, Riverside, CA  92521\\                                                                              and\\
Sydney Meshkov\\
California Institute of Technology, Pasadena, CA 91125
\end{center}
\vspace{1in}

\begin{center}
\textbf{Abstract}            
\end{center}
\begin{quotation}
\noindent We build a model to describe neutrinos based on strict hierarchy, 
incorporating as much as possible, the latest known data, for $\Delta_{sol}$ and $\Delta_{atm}$, and for the mixing angles determined from neutrino oscillation experiments, including that from KamLAND.  Since the hierarchy assumption is  a statement about mass  ratios, it lets us obtain all three neutrino masses. We obtain a mass matrix, $M_\nu$ and a mixing matrix, $U$, where both $M_\nu$ and $U$ are given in terms of powers of $\Lambda$, the analog of the Cabibbo angle $\lambda$ in the Wolfenstein representation, and two parameters, $\rho$ and $\kappa$, each of order one.  The expansion parameter, $\Lambda$, is defined by $\Lambda^2 = m_2/m_3 = \surd (\Delta_{sol}/\Delta_{atm}) \approx$ 0.16,  and $\rho$ expresses our ignorance of the lightest neutrino mass $m_1, (m_1 = \rho  \Lambda^4 m_3$), while  $\kappa$ scales $s_{13}$ to the experimental upper limit, $s_{13}  =  \kappa \Lambda^2 \approx 0.16 \kappa$. These matrices are similar in structure to those for the quark and lepton families, but with $\Lambda$ about $1.6$ times larger than the $\lambda$ for the quarks and charged leptons.  The upper limit for the effective neutrino mass in double $\beta$-decay experiments is $4 \times 10^{-3} eV$ if $s_{13} = 0$ and $6 \times 10^{-3} eV$ if $s_{13}$ is maximal.  The model,  which  is fairly unique, given the hierarchy 
assumption and the data, is compared to supersymmetric extension  and 
texture zero models of mass generation.   
\end{quotation}
\vfil
\newpage

\section{Introduction}

The hierarchical model has been very successful in describing the mass 
patterns and mixing  matrices for quarks and charged leptons~\cite{Fritz}.  Both the mass patterns and mixing angles are  dominated by an expansion 
parameter, which for each family is given by $\lambda = \surd (m_2/m_3)$.  Furthermore the $\lambda$'s for the three families are roughly equal, $0.22<\lambda < 0.25$.  Here, we will try to see  whether neutrinos can be brought simply into the standard fold.  We have a fair handle on the mixing  matrix, but as far as the masses are concerned, we only know the two mass-squared differences, $\Delta_{sol}$  and $\Delta_{atm}$.  This allows the mass ratio $(m_2/m_3)$ to range from about 1 (degeneracy) to small, $\sim 0.1$, (hierarchy). To determine all three masses one more equation is needed  and it is provided by the hierarchy assumption.

The three neutrino mass eigenvalues, $m_1, m_2, m_3$  give a diagonal mass 
matrix.  This matrix can be undiagonalised by the mixing matrix, $U$,  
and the results classified~\cite{Ma} according to possible mass assignments 
consistent with $\Delta_{sol}$ and $\Delta_{atm}$.  In this paper, we try to build a model  based on strict hierarchy, incorporating the known data as much 
as possible.  Since the hierarchy assumption is a statement about mass 
ratios, it lets us obtain all three masses from the two $\Delta$'s.  This 
leads to a mass matrix and mixing matrix, almost entirely in terms of 
powers of $\Lambda$, the analog of the Cabibbo angle $\lambda$ in the Wolfenstein representation.  These matrices are similar in structure to the quark and lepton families, but with $\Lambda$ about $1.6$ times larger than $\lambda$ for the other families.  The effective mass, $<m>$,  measured in $\beta\beta_{0\nu}$ decay~\cite{Pascoli}, is then obtained.  The model,  which  is fairly unique given the hierarchy assumption and the data, is compared to models of mass generation~\cite{Ling,King}.

\section{Determination of the neutrino mass matrix  and mixing matrix}
 
The mixing matrix, $U$,~\cite{Maki} which rotates mass (Majorana) eigenstates $\Psi_{1,2,3}$ into flavor eigenstates $\Psi_{\nu_e,\nu_\mu,\nu_\tau}$ is parameterized as usual
\begin{equation}
U = \left[\begin{array}{ccc}
c_{12} c_{13} & -s_{12} c_{13} & s_{13}e^{-i\partial}\\
s_{12} c_{23} + c_{12} s_{13} s_{23}e^{i\partial} & c_{12} c_{23} - s_{12} s_{13} s_{23}e^{i\partial} & -c_{13} s_{23}\\
s_{12} s_{23} - c_{12} s_{13} c_{23}e^{i\partial} & c_{12} s_{23} + s_{12} s_{13} c_{23}e^{i\partial} & c_{13} c_{23}\end{array}\right]
\end{equation}
Letting $\delta = 0$ (no CP violation), and assuming  maximal mixing for the 
atmospheric oscillation~\cite{Fukuda}, i.e., $s_{23} = c_{23} = 1/\surd 2$, we have for $U$:
\begin{equation}
U = \left[ \begin{array}{ccc}
C c_{13} & -S c_{13} & s_{13}\\
\frac{1}{\sqrt{2}} S + \frac{1}{\sqrt{2}} C s_{13} & \frac{1}{\sqrt{2}} C - \frac{1}{\sqrt{2}} S s_{13} & - \frac{1}{\sqrt{2}} c_{13}\\
\frac{1}{\sqrt{2}} S - \frac{1}{\sqrt{2}} C s_{13} & \frac{1}{\sqrt{2}} C + \frac{1}{\sqrt{2}} S s_{13} & \frac{1}{\sqrt{2}} c_{13}\end{array}\right]
\end{equation}
The angle $\theta_{13}$ is known to be small~\cite{Appol,Boehm,Fogli}, with an upper limit $s_{13} <0.13$ and  at present no lower limit.   We start with the two rotations known not to vanish, $\theta_{12}$ and $\theta_{23}$. With $\theta_{13} = 0$, we have:
\begin{equation}
U = \left[\begin{array}{ccc}
C & -S & 0\\
S/\sqrt{2} & C/\sqrt{2} & -1/\sqrt{2}\\
S/\sqrt{2} & C/\sqrt{2} & 1/\sqrt{2}\end{array} \right]
\end{equation} 
where $S = \sin (\theta_{12}), C = \cos (\theta_{12})$ and we have set the rotation angle $\theta_{23} = \pi/4$ (maximal mixing) and the angle $\theta_{13} = 0$, no CP violation. 

To lowest order in $S$ (expanding $C$ in Eq. (3) in terms of $S$), $U$  is 
given by 
\begin{equation}
U_1 = \left[\begin{array}{ccc}
1 & -S & 0\\
S/\sqrt{2} & 1/\sqrt{2} & -1/\sqrt{2}\\
S/\sqrt{2} & 1/\sqrt{2} & 1/\sqrt{2}\end{array} \right]
\end{equation} 
which, except for the extreme $\theta_{23}$ mixing,  is much like the quark 
mixing matrix. In fact, if we consider $U_1$ to be the result of two 
successive rotations, $U_1 = v_1 v_0$ , we get
\renewcommand{\theequation}{4a}
\begin{equation}
v_0 = \left[\begin{array}{ccc}
1 & 0 & 0\\
0 &  1/\sqrt{2} &  -1/\sqrt{2}\\
0 &  1/\sqrt{2} &  1/\sqrt{2}\end{array} \right] 
\end{equation}
and 
\renewcommand{\theequation}{4b}
\begin{equation}
v_1 = \left[\begin{array}{ccc}
1 & -S/\sqrt{2} & -S/\sqrt{2}\\
S/\sqrt{2} & 1 & 0 \\
S/\sqrt{2} & 0 & 1\end{array} \right] 
\end{equation}
\renewcommand{\theequation}{\arabic{equation}}
\setcounter{equation}{4} 
This suggests that the appropriate expansion parameter for $U$ is given 
by:
\begin{equation}
\varepsilon = S/\surd 2 = \sin (\theta_{12})/\surd 2.
\end{equation}
Recent KamLAND experiments~\cite{Egu} and analyses of this data~\cite{Bahcall,Barger,Pascoli2}, extracting $\theta_{12}$, show that  it is roughly 34$^o$, i.e., $\tan^2 (\theta_{12}) \approx 0.45$, consistent with earlier SNO experiments~\cite{Ahmed}.                                  

Turning to the diagonal mass matrix, we define the conventional 
hierarchical mass pattern by:
\begin{equation}
m_3; m_2: m_1 = 1: \Lambda^2: \rho\Lambda^4.
\end{equation}                                                               
$(\rho = 1$ would correspond to strict hierarchy). The three eigenvalues are then $m_3, m_2 = \Lambda^2 m_3$ and $m_1 = \rho\Lambda^4 m_3$.  

The hierarchy expansion is in terms of the traditional hierarchy parameter, $\Lambda$. 	 \begin{equation}
\Lambda = \surd(m_2/m_3).
\end{equation}
For quarks and charged leptons, the mass matrices, parametrized by $\lambda$, 
and the mixing matrices, given in terms of $\theta_{ij}$ are related.  The 
observed mixing angles of the mixing matrices are given as powers of 
$\lambda$, the Cabibbo angle, as seen in the Wolfenstein~\cite{Wolf} representation of the $V_{CKM}$.  We will try to determine the analogous relationship for neutrinos from the data.  We note that both expansion parameters, $\varepsilon$ and $\Lambda$, can be evaluated, independently, from experimental data. With $\tan^2 (\theta_{12}) = T^2 \approx 0.45$, we have:
\begin{equation}
S^2 \approx 0.31
\end{equation} 
and  therefore
\begin{equation}
\varepsilon^2 = S^2/2 \approx 0.14
\end{equation} 
On the other hand $\Lambda^2 = m_2/m_3$, (the expansion parameter for the mass matrix) can be evaluated in the 
hierarchical expansion, using
\begin{equation}
\Delta_{sol} = m_2^2 - m_1^2 \approx 7.1 \times 10^{-5} eV^2
\end{equation}                      
and  
\begin{equation}
\Delta_{atm} = m_3^2 - m_2^2 \approx 2.7 \times 10^{-3} eV^2
\end{equation}
which forms the ratio
$$
\surd(\Delta_{sol}/\Delta_{atm}) = \surd[(m_2^2  - m_1^2)/(m_3^2 - m_2^2)] $$
\begin{equation}
= \surd[(\Lambda^4 m_3^2 - \rho^2 \Lambda^8 m_3^2)/(m_3^2 - \Lambda^4 m_3^2)].
\end{equation}
The value for $\Delta_{sol}$ is obtained in the analyses of KamLAND data~\cite{Bahcall,Barger}; the value for $\Delta_{atm}$ is obtained from analyses by S. Pascoli, et al.,~\cite{Pascoli2} and G.L. Fogli, et al.,~\cite{Fogli2}. 
Expanding in $\Lambda^2$ we obtain
\begin{equation}
\surd(\Delta_{sol}/\Delta_{atm}) = \Lambda^2 + (1/2)\Lambda^6 (1-\rho^2).
\end{equation} 
Thus to order $\Lambda^4$ we have
\begin{equation}
\Lambda^2 = \surd (\Delta_{sol}/\Delta_{atm}) = 0.16,
\end{equation} 
so that
\begin{equation}
\varepsilon^2 = \Lambda^2 = 0.16
\end{equation}         
or
\begin{equation}
\sin (\theta_{12})/\surd 2 \approx \Lambda
\end{equation} 
 Phenomenologically, at least, there is a close relationship between $\theta_{12}$ and $\Lambda$.

We will use equation 16 as an equality.  We can now 
express both the mass matrix, $M$, as well as the rotation matrix $U$, in 
terms of one parameter, defined in equation 7 as $\Lambda = \surd (m_2/m_3)$,  in 
analogy with the Cabibbo angle for quarks and charged leptons.

The mixing matrix, $U$, is now
\begin{equation}
U=\left[\begin{array}{ccc}
\sqrt{(1-2\Lambda^2)} & - \sqrt{2\Lambda} & 0\\
\Lambda & \sqrt{(1/2)(1-2\Lambda^2)} & -1/\sqrt{2}\\
\Lambda & \sqrt{(1/2)(1-2\Lambda^2)} & -1/\sqrt{2}\end{array}\right]
\end{equation} 
While $\Lambda = 0.4$ is a rather large number, the expansions made in equation 17 are of the square roots and thus, effectively, the expansion 
parameter is $\Lambda^2 = 0.16$.

The mass matrix $M_{\large\nu}$  is given by
\begin{equation}
M_\nu  = U M U^{-1}
\end{equation}
with
\begin{equation}
M = m_3 \left[\begin{array}{ccc}
\Lambda^4 \rho & 0 & 0\\
0 & \Lambda^2 & 0\\
0 & 0 & 1\end{array}\right]
\end{equation}
To order $\Lambda^4, M_\nu$, is given by        
\begin{equation} 
M_{\nu_4} = m_3 \left[\begin{array}{ccc}
\Lambda^4 (\rho+2) & -\Lambda^3 & -\Lambda^3\\
-\Lambda^3 & -\Lambda^4 + (1/2) \Lambda^2 + 1/2 &  -\Lambda^4 + (1/2) \Lambda^2 - 1/2\\
-\Lambda^3 &  -\Lambda^4 + (1/2) \Lambda^2 - 1/2 &  -\Lambda^4 + (1/2) \Lambda^2 + 1/2\end{array}\right]
\end{equation}
where, using eq. (7), $\Lambda = \surd (m_2/m_3) = \surd .16 = 0.4$ and 
\begin{equation}
m_3 = \surd (\Delta_{atm} + \Delta_{sol}) = 5.2 \times 10^{-2} eV. 
\end{equation}
While the  analytic expressions are to the orders in $\Lambda$ indicated, the 
data, unfortunately, are not.  Thus, all results will be given to at most two significant figures.

We now can determine the masses, $m_1, m_2$ and $m_3$ to order $\Lambda^4$:
\begin{equation}
m_3 = \surd (\Delta_{atm} + \Delta_{sol}) = 5.2\times 10^{-2} eV
\end{equation} 
\begin{equation}
m_2 = \Lambda^2  m_3 = \surd (\Delta_{sol}/\Delta_{atm}) m_3 = 8.3 \times 10^{-3} eV 
\end{equation}                                                    
and
\begin{equation} 
m_1 = \rho\Lambda^4 m_3 = \rho (\Delta_{sol}/\Delta_{atm}) m_3 = 1.3 \rho \times 10^{-3} eV
\end{equation}
We may let $\rho$ range from, say, $2$ to $-2$ and still consider $|m_1|$  to be of order $\Lambda^4$ or smaller.  These masses follow directly from the hierarchy assumption and the experimental values: $\Delta_{sol} = m_2^2 - m_1^2 \approx 7.1 \times 10^{-5} eV^2$ and $\Delta_{atm} = m_3^2 - m_2^2 \approx 2.7 \times 10^{-3} eV^2$ and are independent of the mixing matrix $U$. The effective neutrino mass $<m>$   measured in $\beta\beta_{0\nu}$ decay,~\cite{Pascoli} however, does depend on the neutrino mass matrix, $M_\nu$ and  is  given by $M_{\nu_{ee}}$, the $1,1$ matrix element.  From equations (17) and (19) we have, to order $\Lambda^6$
\begin{equation}
<m> = M_{\nu_{ee}} = m_3 \Lambda^4 [\rho (1-2\Lambda^2) + 2] = (0.7 \rho +2) \times 1.3  \times 10^{-3} eV 
\end{equation} 
Taking $\rho=2, (m_1 = 2.6 \times 10^{-3} eV)$ as an extreme case, we obtain as  an upper limit (with $s_{13} = 0$):
\begin{equation}
<m>\quad \leq 4 \times 10^{-3} eV
\end{equation}                             
On the other hand, $m_1$ and therefore $\rho$, may be negative. For $\rho = - 2, (m_1 = - 2.6 \times 10^{-3} eV)$, we have 
\begin{equation}
\langle m\rangle = M_{\nu_{ee}} = \Lambda^6 {m_{3}} = 8 \times 10^{-4} eV. 
\end{equation} 
Thus, the limits on $<m>$ in this model are, approximately,
\begin{equation}
10^{-3} < \quad <m>\quad < 4\times 10^{-3} eV
\end{equation}  

\section{Inclusion of $s_{13} = \sin (\theta_{13})$}

While the present data do not demand a non-vanishing $\theta_{13}$, several 
models of hierarchy generation do~\cite{Ma,Giunti}.   We want to investigate the effect of a finite $s_{13}$.

Since we know that $s_{13} < 0.13$~\cite{Appol,Boehm,Fogli} and  there is  at present no lower limit, we will scale $s_{13}$:
\begin{equation}
s_{13} = \kappa \Lambda^2 
\end{equation}                                                                      where $1 \geq |\kappa| \geq 0$.

Substituting  for $s_{13}$ in eq. (20) and forming the mass matrix $M_\nu$, 
we  obtain to order $\Lambda^4$:
\begin{equation}
M_{\nu_4} = m_3 \left[\begin{array}{ccc}
\Lambda^4 (\kappa^2 + \rho + 2) & -\Lambda^3 - \frac{1}{\sqrt{2}}\Lambda^2 \kappa &  -\Lambda^3 + \frac{1}{\sqrt{2}} \Lambda^2 \kappa\\
-\Lambda^3 - \frac{1}{\sqrt{2}} \Lambda^2 \kappa & -\frac{1}{2} \Lambda^4 (\kappa^2 + 1) + \frac{1}{2} \Lambda^2 + \frac{1}{2} & \frac{1}{2} \Lambda^4 (\kappa^2 + 1) + \frac{1}{2} \Lambda^2 - \frac{1}{2}\\
-\Lambda^3 + \frac{1}{\sqrt{2}} \Lambda^2 \kappa & \frac{1}{2} \Lambda^4 (\kappa^2 - 1) + \frac{1}{2} \Lambda^2 - \frac{1}{2} & -\frac{1}{2} \Lambda^4 (\kappa^2 + 1) + \frac{1}{2} \Lambda^2 + \frac{1}{2}\end{array}\right]
\end{equation}  
We now have two parameters, $\rho$ and $\kappa$, where $\rho$ is defined by $m_1 = \rho \Lambda^4 m_3$ and \linebreak $s_{13} = \kappa \Lambda^{2}, |\kappa| \leq 1$.  There are three special regimes for $\kappa$ which are interesting.

\begin{enumerate}
\item $\kappa = 0, s_{13} = 0$   
This is the case which was discussed earlier. We give here the leading 
elements of $M_\nu$ which depend on $\kappa$. 
\begin{equation}
M_{\nu_{ee}} = <m> = \Lambda^4 (\rho + 2) m_3
\end{equation}
\begin{equation}
M_{\nu_{e\mu}} = M_{\nu_{e3}} = - \Lambda^3 m_3
\end{equation}
\item $\kappa \approx 1$ (upper limit), $s_{13} = \Lambda^2$  
\begin{equation}
M_{\nu_{ee}} = <m> = \Lambda^4 (\rho +2 + \kappa^2)m_3 = \Lambda^4 (\rho +3) m_3 
\end{equation}
\begin{equation}
M_{\nu_{e\mu}} = - M_{\nu_{e\tau}} =\pm (1/\surd 2) \Lambda^2 m_3
\end{equation}
 and the most interesting possibility,
\item $\kappa = \kappa' \Lambda$ i.e., $s_{13} = \kappa' \Lambda^3$.  For case (3)  the mass matrix to order $\Lambda^4$ is:
\begin{equation}
M_{\nu_4} = m_3\left[\begin{array}{ccc}
\Lambda^4 (\rho +2) & -\Lambda^3 \left(1 + \frac{\kappa'}{\sqrt{2}}\right) & - \Lambda^3 \left( 1 - \frac{\kappa'}{\sqrt{2}}\right)\\
-\Lambda^3 \left( 1 + \frac{\kappa'}{\sqrt{2}}\right) & - \Lambda^4 + \Lambda^2/2 + \frac{1}{2} & - \Lambda^4 + \Lambda^2/2 - \frac{1}{2}\\
-\Lambda^3 \left( 1 - \frac{\kappa'}{\sqrt{2}}\right) & - \Lambda^4 + \Lambda^2/2 - \frac{1}{2} & - \Lambda^4 + \Lambda^2/2 + \frac{1}{2}\end{array}\right]
\end{equation}                       
\end{enumerate}
and
\begin{equation}
M_{\nu_{e\mu}} = - \Lambda^3 [1 + (1/\surd 2) \kappa'] m_3
\end{equation}
\begin{equation}
M_{\nu_{e\tau}} = - \Lambda^3 [1 - (1/\surd 2) \kappa'] m_3
\end{equation}
Case (3) is the only case which allows a zero in an off diagonal 
element. Texture zeros have been considered as a possible source of 
hierarchies and mixing angles~\cite{Ling,King,Framp,Desai}.  From Eq. (35), we see that only $\kappa = \kappa' \Lambda$ provides the possibility of having two texture zeroes.

Taking $\rho = - 2$ and $\kappa' = \pm \surd 2$  will make $M_{\nu_{ee}}$  and either $M_{\nu_{e\mu}}$ or $M_{\nu_{e\tau}}$ vanish to order $\Lambda^4$.  With $\kappa' = - \surd 2, M_{\nu_4}$   becomes
\begin{equation}
M_{\nu_4} = m_3\left[\begin{array}{ccc}
0 & 0 & -2\Lambda^3\\
0 & -\Lambda^4 + \Lambda^2/2 + \frac{1}{2} & -\Lambda^4 + \Lambda^2/2 - \frac{1}{2}\\
-2\Lambda^3 & -\Lambda^4 + \Lambda^2/2 - \frac{1}{2} & -\Lambda^4 + \Lambda^2/2 + \frac{1}{2}\end{array}\right]
\end{equation}     
In terms of masses we substitute $m_2 = \Lambda^2 m_3, m_1 = \rho \Lambda^4 m_3 = - 2 \Lambda^4 m_3$ and get  
\begin{equation}
M_{\nu_4} = \left[\begin{array}{ccc}
0 & 0 & -\sqrt{(-2m_1m_2)}\\
0 & \frac{1}{2} (m_1 + m_2 + m_3) & \frac{1}{2} (m_1 + m_2 - m_3)\\
-\sqrt{(-2m_1 m_2)} & \frac{1}{2} (m_1 + m_2 - m_3) & \frac{1}{2} (m_1 + m_2 + m_3)\end{array}\right]
\end{equation}
Eq. (39) is identical to the matrix derived  for the hierarchical 
case  by B. R. Desai et. al.\cite{Desai}, who systematically categorize the 
neutrino mass matrices, consistent with experimental constraints, with 
two texture zeros.  Note that for this model $<m> = M_{\nu_{ee}}$ vanishes to order $\Lambda^4$.
In order to compare with a recent model for hierarchy generation,\cite{Ling} 
we continue with $\rho = - 2$, but do not specify $~\kappa'$ in $s_{13} = \kappa' \Lambda^3$.  In that case , keeping only the leading order in  $\Lambda$ in each matrix element of $M_{\nu_6}$, we obtain :
\begin{equation}
M_{\nu_6} = m_3\left[\begin{array}{ccc}
\Lambda^6 (4 + \kappa^{\prime 2}) & - \Lambda^3 (1 + \frac{1}{\sqrt{2}} \kappa') & - \Lambda^3 (1 - \frac{1}{\sqrt{2}} \kappa')\\
- \Lambda^3 (1 + \frac{1}{\sqrt{2}} \kappa') &~~ \frac{1}{2} & - \frac{1}{2}\\
- \Lambda^3 (1 - \frac{1}{\sqrt{2}} \kappa') & -\frac{1}{2} &~~ \frac{1}{2} \end{array}\right]
\end{equation}              
The leading orders of $\Lambda$ in each matrix element are the orders 
indicated in the work of Ramond, et al.,\cite{Ling} (and `tuned' by Fishbane and Kaus~\cite{Fish}).  This model suggests , within a super symmetric extension of the standard model, that the existence of mass hierarchies within 
fermionic sectors imply at least one additional $U(1)$ family symmetry 
one of which must be anomalous, with a cancellation of its anomaly 
through the Green-Schwarz mechanism then implying relations  across 
fermionic sectors.  This has the additional  property of  predicting 
$\Lambda$,  which should be the same for all families.  However, the data for 
neutrinos, $\Delta_{atm}$ and $\Delta_{sol}$, suggest that $\Lambda \approx 0.4$, while for the other family sectors, we have the traditional $\lambda \approx 0.25$.  
\section{Summary}

We have shown that the assumed hierarchy pattern and the present data 
imply that a mixing matrix, $U$, and mass matrix, $M_\nu$ may be expressed in 
terms of powers of the expansion parameter $\Lambda$ and two parameters $\rho$ and $\kappa$ of order one.  The parameter $\rho$ expresses our ignorance of the lightest neutrino mass, $m_1$, where $m_1 = \rho  \Lambda^4  m_3 \approx 1.3 \rho \times 10^{-3} eV$ and  $\kappa$ scales $s_{13}$ to the experimental upper limit, $s_{13} = \kappa \Lambda^2 \approx 0.16 \kappa$.  The simplicity of $U$ and $M_\nu$ comes from the observed relationship, $S^2/2 = \Lambda^2 = m_2/m_3$.  The expansion parameter, $\Lambda$, where $\Lambda^2 = m_2/m_3 = \surd (\Delta_{sol}/\Delta_{atm} \approx 0.16$ is 
identical in spirit, though not in value, to the Wolfenstein parameter
~\cite{Wolf}, $\lambda$, in the quark $V_{CKM}$ and is measured by solar and atmospheric oscillation experiments.   The upper limit for the effective neutrino mass in double $\beta$-decay experiments is $4 \times 10^{-3} eV$ if  $s_{13} = 0$ and $6 \times 10^{-3} eV$ if $s_{13}$ is maximal.  

The models of hierarchical mass generation that we compared to, 
supersymmetric extension~\cite{Ling} and texture zeroes~\cite{Ling,King,Framp,Desai} each demand that $M_{\nu_{ee}}$ be of order $\Lambda^6$ or even vanish.  This implies $\rho = - 2, m_1 \approx -2.6 \times 10^{-3} eV$, in Eq. (30), and thus $<m> \approx 10^{-3} eV$ or smaller.  Both of these models require the $M_{\nu_{e\mu}}$ and $M_{\nu_{e\tau}}$ terms to be of order $\Lambda^3$ or smaller. Therefore $s_{13}$ in Eq. (30) has to be of order $\Lambda^3$, i.e.,  $s_{13} = \kappa' \Lambda^3 = 0.06 \kappa'$, where $\kappa'$ is of order unity or smaller.  More specifically, for a texture zero in $M_{\nu_{e\mu}}$ or $M_{\nu_{e\tau}}$  one must have $\kappa' = \pm \surd 2$ or $s_{13} \approx 0.09$.  All these demands are well 
below the present upper experimental upper limits of $<m>$ and $s_{13}$.  

We thank the Aspen Center for Physics, where this work started, for 
its hospitality and stimulating atmosphere.

 \end{document}